\documentclass[aps,prl,notitlepage,twocolumn,10pt,a4paper]{revtex4-1}

\usepackage{xcolor,graphicx,ulem,soul}
\usepackage{amsmath,amssymb}
\usepackage[colorlinks=true,urlcolor=blue,citecolor=blue,linkcolor=blue]{hyperref}
\usepackage{braket}
\usepackage{cancel}
\usepackage{comment}
\usepackage[utf8]{inputenc}

\newcommand{\w}{\omega}
\newcommand{\W}{\Omega}
\newcommand{\cdag}{\hat{c}^{\dagger}}
\newcommand{\dddag}{\hat{d}^{\dagger}}
\newcommand{\xdag}{\hat{x}^{\dagger}}
\newcommand{\ydag}{\hat{y}^{\dagger}}
\newcommand{\adag}{\hat{a}^{\dagger}}
\newcommand{\bdag}{\hat{b}^{\dagger}}
\newcommand{\vac}{\ket{\text{vac}}}
\newcommand{\ud}{\textrm{d}}
\newcommand{\modsqr}[1]{\left| #1 \right|^2}

\begin{document}


\title{Heralding multiple photonic pulsed Bell-pairs via frequency-resolved entanglement swapping}


\author{Sofiane Merkouche$^{*1}$, Val\'{e}rian Thiel$^1$, Alex O.C. Davis$^{2}$, and Brian J. Smith$^{1}$}
\email[]{Corresponding author: sofianem@uoregon.edu}
\affiliation{$^1$ Department of Physics and Oregon Center for Optical, Molecular, and Quantum Science, University of Oregon, Eugene, Oregon 97403, USA\\
$^2$Centre for Photonics and Photonic Materials, Department of Physics, University of Bath, Bath, BA2 7AY, UK}


\date{\today}

\begin{abstract}
Entanglement is a unique property of quantum systems and an essential resource for many quantum technologies. The ability to transfer or swap entanglement between systems is an important protocol in quantum information science. Entanglement between photons forms the basis of distributed quantum networks and the demonstration of photonic entanglement swapping is essential for their realization. Here an experiment demonstrating entanglement swapping from two independent multimode time-frequency entangled sources is presented, resulting in multiple heralded temporal-mode Bell states. Entanglement in the heralded states is verified by measuring conditional anti-correlated joint spectra as well as quantum beating in two-photon interference. Our proof-of-concept experiment is able to distinguish up to five orthogonal Bell pairs within the same setup, limited in principle only by the entanglement of the sources.
\end{abstract}


\maketitle


\paragraph*{Introduction.---}\hspace{-3ex}
Entanglement, the correlations displayed between subsystems of a multipartite quantum system, is one of the most distinguishing properties of quantum theory and a significant resource for quantum information science (QIS). Entanglement swapping \cite{Zukowski1993} is a protocol that enables entanglement of quantum systems that have never interacted and can be separated by large distances \cite{Riedmatten2005} or even time \cite{Peres2000} \cite{Ma2012}.  This protocol underpins efforts to realize large-scale quantum networks as the core element of quantum repeaters \cite{Briegel1998}, in addition to shedding light on the fundamental nature and extent of non-locality of multipartite quantum systems. 

Entanglement swapping relies on the ability to perform projective measurements onto entangled states. For photonic states, projective measurements onto two-photon entangled states can be implemented using a beam splitter and mode-resolved measurements. Entanglement swapping has been experimentally demonstrated using photons entangled in their polarization \cite{Pan1998}, spatial \cite{Zhang2017}, and temporal \cite{Halder2007} degrees of freedom. Recent efforts have shown that temporal- or pulsed-mode encoding offers unique opportunities for QIS \cite{Brecht2015}. Thus addressing pulse-mode entanglement manipulation and verification is a timely topic \cite{Graffitti2020}.

In this Letter, we report an experiment demonstrating the swapping of time-frequency two-photon entanglement between two independent multimode photon pair sources. This is enabled by multiplexed frequency-resolved detection \cite{Davis2016} to implement projective measurements of multiple temporal-mode Bell states. The entanglement of the heralded two-photon states is verified by measurement of two-photon quantum beats \cite{Ou1988}, and the joint-spectral intensity arising from four-fold frequency resolved measurements. To the best of our knowledge, this is the first experiment to demonstrate both heralding and discrimination of multiple Bell pairs with a single source and measurement apparatus, as well as the first experiment employing simultaneous time-of-flight spectrometry of four photons.

\begin{figure*}[t]
	\centering
	\includegraphics[width=.85\linewidth]{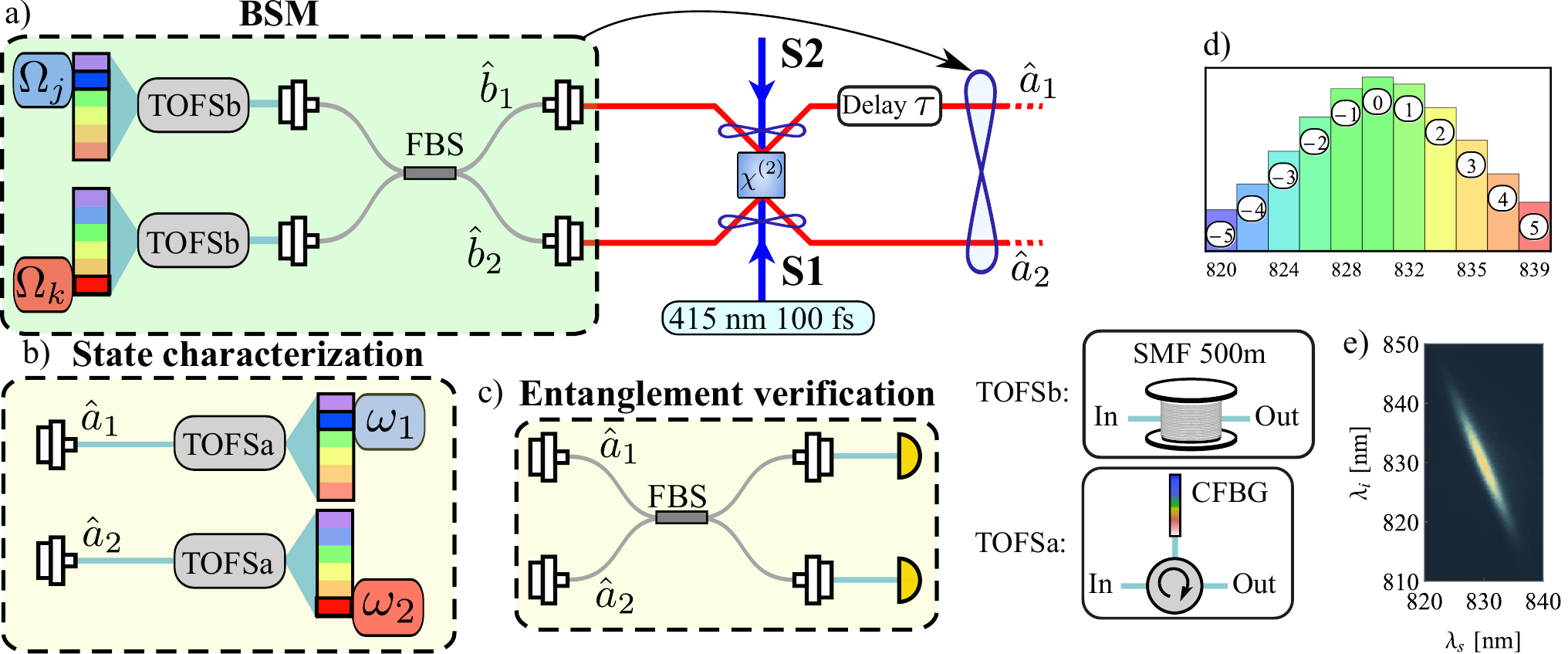}
	\caption{a) Experimental setup - see main text for description. BSM: frequency-resolved Bell-state measurement implemented on the idler photons. FBS: fiber beamsplitter. SMF: single mode fiber. CFBG: chirped fiber Bragg grating. b) State characterization  - joint spectral measurement of the signal photons, conditioned on the BSM. c) Entanglement verification - two-photon interference as a function of relative delay $\tau$, conditioned on the BSM. TOFSa and TOFSb are disperion-based time-of-flight spectrometers. d) Labeling convention for $\Omega_{j(k)}$ measurements, (e.g. $\Omega_0$ corresponds to a bin centered at 830 nm). e) Measured JSI for each of the sources, where $\lambda_s(\lambda_i)$ is the signal (idler) wavelength.}	 \label{fig:setup}
\end{figure*}

\textit{Theory --}
The two-photon term of output state of a single spontaneous parametric down conversion (SPDC) source can be expressed as

\begin{equation}
	\ket{\psi}=\int \ud\omega_s \ud\omega_i f(\omega_s, \omega_i) \adag(\omega_s) \bdag(\omega_i)\ket{\mathrm{vac}}
\end{equation}
where $\adag(\omega)\ (\bdag(\omega))$ creates a photon with frequency ${\omega}$ in the signal (idler) mode. The function $f(\omega_s,\omega_i)$ is the normalized complex joint spectral amplitude (JSA), and its modulus squared, $|f(\omega_s,\omega_i)|^2$, is the joint spectral intensity (JSI). The JSI is the two-photon probability density function in frequency space. The state contains spectral entanglement when the JSA is not factorable; that is, when $f(\omega_s,\omega_i) \neq f_s(\omega_s)f_i(\omega_i)$. We assume that the process is single-mode in the polarization and transverse spatial degrees of freedom, so that only the time-frequency degrees of freedom are relevant.


Our experiment makes use of two independent, identical SPDC sources, producing the state

\begin{equation}
	\begin{gathered}
		\ket{\psi_{12}}=\int \ud\omega_s\ud\omega_i\ud\omega_s'\ud\omega_i'  f(\omega_s,\omega_i)\hat{a}_1^{\dagger}(\omega_s)\hat{b}_1^{\dagger}(\omega_i)\\ \times f(\omega'_s,\omega'_i)\hat{a}_2^{\dagger}(\omega'_s)\hat{b}_2^{\dagger}(\omega'_i)\ket{\mathrm{vac}}
	\end{gathered}
\end{equation}
where the 1 and 2 subscripts denote the first and second sources, respectively. Entanglement swapping requires performing a partial Bell-state measurement (BSM) on the idler fields $b_1$ and $b_2$, which is achieved by interfering the fields at a 50:50 beamsplitter and performing a frequency-resolved coincident detection at the output, at frequencies $\Omega_j$ and $\Omega_k$. This measurement projects the input idler fields onto the two-color singlet Bell state $\ket{\psi^-_{jk}}=\frac{1}{\sqrt{2}} (\ket{\Omega_j}_{b1}\ket{\Omega_k}_{b2}-\ket{\Omega_k}_{b1}\ket{\Omega_j}_{b2})$, and the state heralded in the signal fields is well-approximated 
by (see Supplemental)

\begin{equation}
	\begin{gathered}
		\ket{\Psi^-_{jk}}=\frac{1}{\sqrt{2\mathcal{C}_{jk}}}\left(\ket{\phi_j}_1\ket{\phi_k}_2 - \ket{\phi_k}_1\ket{\phi_j}_2\right).
	\label{eq:state}
	\end{gathered}
\end{equation}
Here we have defined
\begin{equation}
	\begin{gathered}
		\ket{\phi_{j(k)}}_{1(2)} = \int \ud\omega \phi_{j(k)}(\omega) \adag_{1(2)}(\omega)\vac, 
	\end{gathered}
\end{equation}
where $\phi_{j(k)} (\omega) \propto f( \omega,\Omega_{j(k)})$ are normalized Gaussian amplitude functions given by $\phi_{j(k)}(\omega) \propto \exp{[-(\omega-\omega_{j(k)})^2/4\sigma^2]}$, with $\omega_{j(k)}$ and $\sigma$ determined by the JSA, and $\mathcal{C}_{jk} = 1-|\braket{\phi_j|\phi_k}|^2$  (see Supplemental). 


\begin{figure*}[t]
	\centering
	\includegraphics[width=\linewidth]{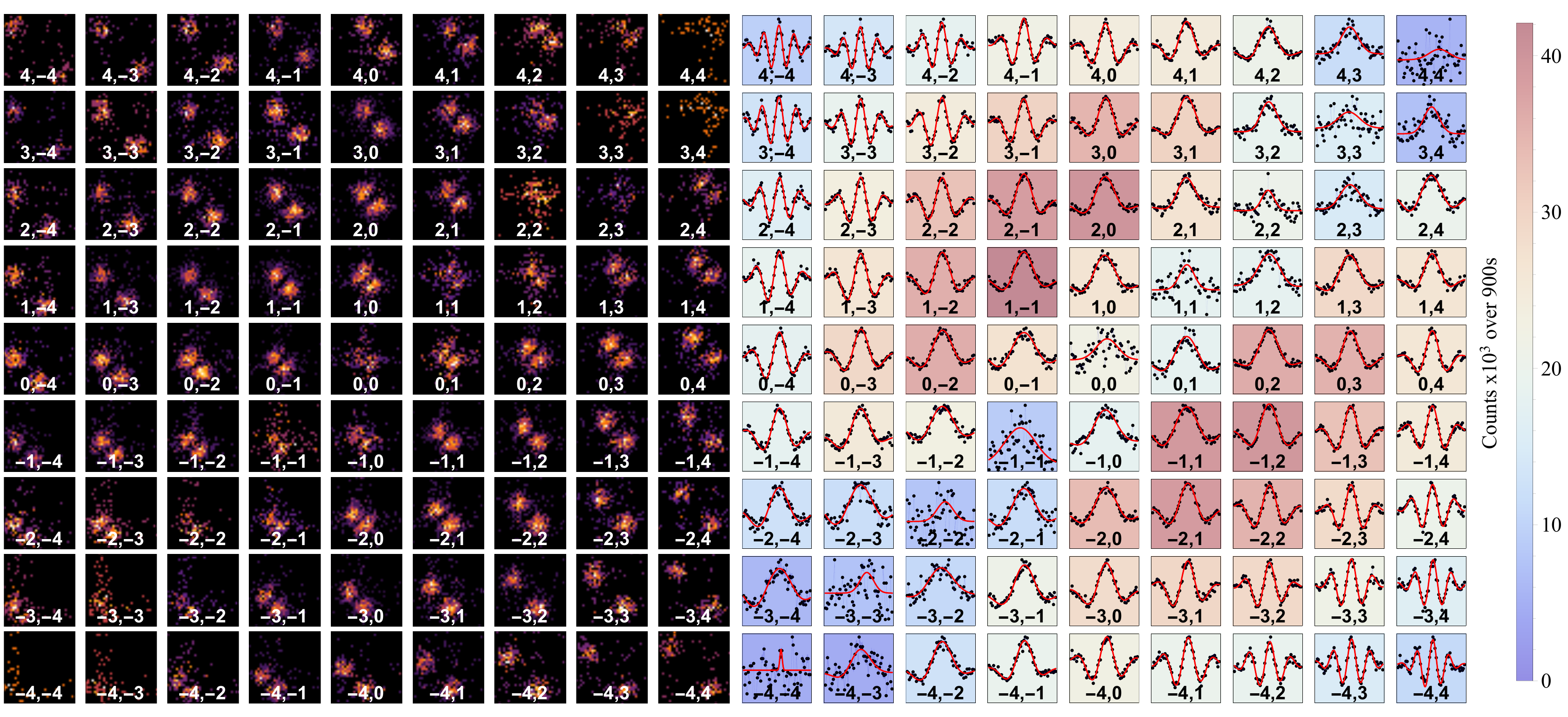}
	\caption{Left, state characterization: array of the measured JSI's $F_{jk}$ of the signal photons, conditioned on the $(\Omega_j,\Omega_k)$ outcomes of the BSM. Right, entanglement verification: array of interference fringes $P_{jk}(\tau)$, conditioned on the same $(\Omega_j,\Omega_k)$ outcomes, verifying entanglement of the states. The background color indicates the total number of counts for each plot, and thus the entire array maps out the $p_{jk}$ matrix. Each array was obtained in a single measurement run.}\label{fig:fringes}
\end{figure*}
The heralded state $\ket{\Psi^-_{jk}}$ can be characterized by measuring its JSI, which is given by

\begin{equation}
	F_{jk}(\omega_1,\omega_2) = \frac{1}{2\mathcal{C}_{jk}}\left|\phi_j(\omega_1)\phi_k(\omega_2)-\phi_k(\omega_1)\phi_j(\omega_2)\right|^2.
\end{equation}
In order to verify entanglement in the state $\ket{\Psi^-_{jk}}$, beyond classical correlations, two-photon interference is used in a manner similar to the method employed in reference \cite{Graffitti2020}. Here the heralded signal photons are detected in coincidence at the output of a 50:50 beamsplitter, as a function of relative arrival time delay $\tau$. For the input state $\ket{\Psi^-_{jk}}$, the coincidence probability is given by (see Supplemental)

\begin{equation}
	P_{jk}(\tau)=\frac{1}{2} + \frac{1}{2}e^{-\tau^2 \sigma^2}\cos\left[(\omega_j - \omega_k)\tau\right],
	\label{eq:pjkt}
\end{equation}
which oscillates at the difference frequency $\omega_j-\omega_k$. These oscillations, obtained without filtering of the interfering fields, are a hallmark of two-color entanglement of the input state (see, for instance, reference \cite{Ramelow2009}).


If the heralding idler photons are not resolved in the $(\Omega_j,\Omega_k)$ space, the signal photons are heralded in the state

\begin{equation}
	\hat{\rho}=\sum_{j,k} p_{jk}\ket{\Psi^-_{jk}}\bra{\Psi^-_{jk}},
	\label{eq:rho}
\end{equation}
where $p_{jk}$, normalized as $\sum_{j,k} p_{jk} = 1$, is the joint spectral distribution of the idler photons \textit{at the output} of the beamsplitter, and gives the probability of heralding the state $\ket{\Psi^-_{jk}}$. Likewise, the JSI of this state will be given by

\begin{equation}
	F(\omega_1,\omega_2)= \sum_{j,k} p_{jk} F_{jk}(\omega_1,\omega_2).
	\label{eq:bigf}
\end{equation}
The state $\hat{\rho}$ is a mixed state which retains the antisymmetry of its constituent states $\ket{\Psi^-_{jk}}$. This is evidenced by its two-photon interference pattern, given by

\begin{equation}
	P(\tau)=\sum_{j,k} p_{jk} P_{jk}(\tau).
	\label{eq:pt}
\end{equation}
where, notably, we still expect a coincidence peak to survive at $\tau=0$. That is, the antisymmetry of the state $\hat{\rho}$ is the basis for this predicted antibunching \cite{Fedrizzi2009}.

\paragraph*{Experiment.---}\hspace{-3ex}
Our experimental setup is shown schematically in Fig. \ref{fig:setup}.  For the light source we use ultrashort (100 fs) pulses from a titanium-doped sapphire (Ti:Sapph) laser oscillator at a central wavelength of 830 nm and a repetition rate of 80 MHz. These pulses are frequency-doubled in a 1 mm-long birefringent $\mathrm{BiB_3O_6}$ crystal (BiBO) to generate the blue (415 nm) pump for the SPDC sources. SPDC occurs at a second, 2.5 mm-long BiBO, which is double-passed to generate a pair of frequency-entangled photons on the first pass (source 1), and on the second pass (source 2). This double-pass configuration ensures that the two sources are identical. Type II phase matching permits the deterministic separation of the signal and idler photons using polarizing beamsplitters, after the blue pump has been filtered out using dichroic mirrors. Signal and idler photons from both sources are collected into polarization-maintaining single-mode fibers (PM fibers) and directed to the remainder of the set-up for analysis and entanglement swapping. We measure a pair detection rate of up to 300 kHz from each source using superconducting nanowire single-photon detectors (SNSPDs) from IDQuantique.
\begin{figure*}[t]	 
	\centering
	\includegraphics[width=\linewidth]{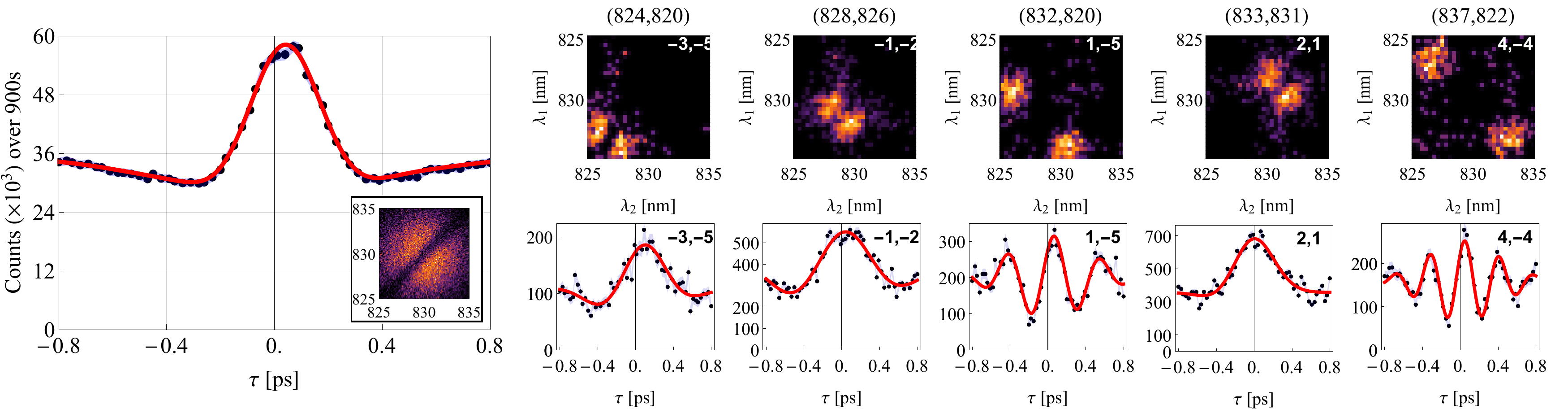}
	\caption{Left: The integrated interference fringes $P(\tau)$ as given by Eq. \ref{eq:pt}. The red plot is not a fit, but the sum of the fits from Fig. \ref{fig:fringes}b. The inset is the integrated JSI $F(\omega_1,\omega_2)$ as given by Eq. \ref{eq:bigf}. On the right is a set of JSI's and interference fringes corresponding to 5 quasi-orthogonal modes, satisfying an overlap of $\leq 0.15$ as described in the text, chosen from the full set  from Fig. \ref{fig:fringes}.}\label{fig:peak_and_ortho_modes}
\end{figure*}  
The joint spectral intensity of each source is measured efficiently using a time-of-flight spectrometer consisting of a pair of 500 m-long fibers (TOFSb). Each photon from the signal-idler pair is passed through the dispersive fiber, imparting a wavelength-dependent delay relative to the Ti:Sapph reference pulse train. Time-resolved coincidence detections at the output, using a time-to-digital converter (ID900) with a resolution of about 30 ps, provide a direct measure of the joint spectral intensity with a resolution of about 0.5 nm (see Fig. \ref{fig:setup}e). Assuming negligible phase correlations \cite{Davis2020}, we estimate the amount of entanglement in the state by taking the square root of the JSI and calculating the Schmidt number \cite{Law2000}, for which we obtain a value of $K \sim 4$. 
        
The entangling BSM is performed on the idler photons by routing them to a polarization-maintaining fiber beamsplitter (FBS) to ensure mode-matching and indistinguishability at the output, while temporal matching is achieved using a free-space delay line. To this end, a delay line is scanned in the path of the idler photon from source 1, while monitoring coincidences between the two outputs of the FBS until a Hong-Ou-Mandel dip is observed. To complete the BSM, the idler photons after the FBS are routed through the TOFSb, and detected with a resolution of about 1.5 nm. Each coincident idler photon detection at the frequency pair $(\Omega_j,\Omega_k)$ heralds a distinct state $\ket{\Psi^-_{jk}}$, as defined in Eq. \ref{eq:state}, in the signal photons.


We characterize the heralded state $\ket{\Psi^-_{jk}}$ by measuring its JSI using a similar set-up to TOFSb. Here, a chirped fiber Bragg gratings (TOFSa) are used instead of the long fibers, imparting a large dispersion and giving a high spectral resolution (0.1 nm, vs 1.5 nm for TOFSb), but with higher losses (10 dB, vs 1 dB for TOFSb) \cite{Davis2016}. The BSM on the idler photons associates with each $(\Omega_j,\Omega_k)$ pair a distinct signal-pair JSI $F_{jk}(\omega_1,\omega_2)$, corresponding to the state $\ket{\Psi^-_{jk}}$. In Fig. \ref{fig:fringes} a) we display an array of the measured JSI's, with $j,k \in [-4,4]$ according to the convention in Fig. \ref{fig:setup}. This data was all obtained in a single measurement run.

To perform the entanglement verification, the signal photons are routed through another 50:50 FBS, and detected in coincidence at the output while scanning a free-space time delay $\tau$ in the arm of the signal from source 1. As for the idler case, delay matching was obtained by measuring the unheralded signal coincidences and observing a Hong-Ou-Mandel dip. For each of the heralded states $\ket{\Psi^-_{jk}}$, we observe coincidence fringes that oscillate at the angular frequency difference $(\omega_j-\omega_k)$, as predicted in equation \ref{eq:pjkt}. We plot the corresponding array of these results in Fig. \ref{fig:fringes} b. Because we use probabilistic sources, it should be recalled that the probability of one of the sources firing two pairs of photons is on the same order as the probability of each source firing a single pair. This contributes to additional terms in the heralded state which are inherent to all similar entanglement swapping experiments \cite{Wagenknecht2010}, and which appear as a constant background that we measure and subtract in the entanglement verification measurements \cite{Graffitti2020}. After this background subtraction, the measured visibility of the interference fringes is roughly 75\%, consistent with the maximum visibility expected from source matching measurements (see Supplemental). Finally, it is notable that the total number of counts measured in each $(j,k)$ bin of the arrays in figure \ref{fig:fringes} is in fact a measure of $p_{jk}$. We highlight this by coloring the background of the interference plots as a function of the total number of counts, such that the entire array can be seen as a plot of the JSI of the idler photons after the beamsplitter. Note the ridge along the diagonal $(j=k)$ due to Hong-Ou-Mandel interference.

As pointed out in the previous section, performing an unresolved BSM on the idler photons, where the measurement is integrated over all $(\Omega_j,\Omega_k)$, projects the signal photons onto the mixed state $\rho$ denoted in Eq.\ref{eq:rho}. Although mixed, this state is a convex combination of antisymmetric Bell pairs over frequency space. Two notable features arise from this fact. First, there are no coincidences where $\omega_j=\omega_k$, due to the aforementioned Hong-Ou-Mandel interference in the idler BSM, and this contributes to a ridge along the diagonal of the integrated JSI $F(\omega_1,\omega_2)$, displayed as an inset in Fig. \ref{fig:peak_and_ortho_modes}a. Second, the antisymmetry of the state is preserved, and this is evidenced by the fully visible peak in the integrated two-photon interference scan of $P(\tau)$, which is displayed in the main plot of Fig. \ref{fig:peak_and_ortho_modes}a. Note also that the red curve in that plot is not a new fit to the data, but rather just the sum of the individual fits to the $P_{jk}(\tau)$. Taken all together, our data nicely highlights the quantum nature of measurement, whereby different quantum states arise as a consequence of different measurement results.

As a final point, we note that although the number of states $\ket{\Psi^-_{jk}}$ that we can resolve is essentially limited by the resolution of the spectrometers on the heralding side, not all these states will be orthogonal, because the number of available orthogonal modes in the sources is finite to begin with. This is indeed the property that is quantified by the Schmidt number. A combinatorics argument shows that, for two identical sources with Schmidt number $K$, one can herald at most $K(K-1)/2$ orthogonal Bell pairs \cite{Zhang2017}. The ideal Bell state measurement for our scheme would resolve the idlers in the Schmidt mode basis of the sources (by using a quantum pulse gate, for instance \cite{Reddy2018}), thus automatically heralding the signals in orthogonal Bell pairs. Since we instead resolve the idlers into frequency bins, we herald more signal states than those comprising an orthogonal set. We can then simply choose a quasi-orthogonal set of states from our data which satisfy an overlap criterion of $\int \ud \omega_1 \ud \omega_2 F_{jk}(\omega_1,\omega_2) F_{j'k'}(\omega_1,\omega_2) \leq \epsilon$, $\forall(j, k) \neq (j', k')$, where $\epsilon$ can be chosen arbitrarily small. In an application setting such as multiplexed entanglement distribution, one could in principle restrict consideration to this smaller set of $j,k$ heralding events which correspond to orthogonal states, in order to avoid any unwanted crosstalk between frequency channels. In Fig. \ref{fig:peak_and_ortho_modes}b, we show a representative set of quasi-orthogonal modes selected with $\epsilon = 0.15$. In this case there are 5 orthogonal modes, while our estimated $K$ of 4 predicts a maximum of 6 orthogonal modes.

\begin{figure}[t]
	\centering
	\includegraphics[width=.5\linewidth]{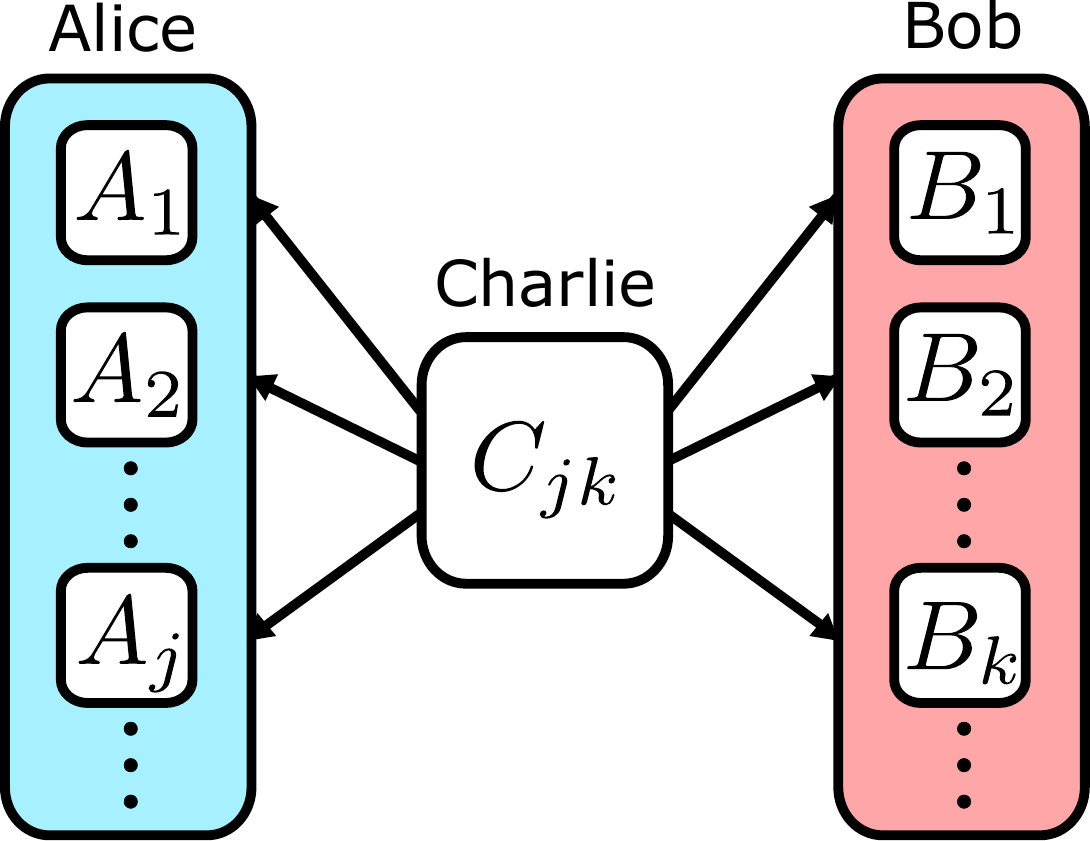}
	\caption{Frequency-multiplexed entanglement swapping scheme. Charlie performs a frequency-resolved BSM on photons from sources with high-dimensional entanglement. Each independent result $C_{jk}$ enables distribution of a photonic Bell pair to a separate pair of parties $A_j$ and $B_k$ on the Alice and Bob sides, even if using a single fiber transmission line.} 
	\label{fig:quantinder}
\end{figure}

\paragraph*{Discussion.---}\hspace{-3ex}
One conceivable application of this work is a multi-party quantum key distribution network, as depicted schematically in Fig. \ref{fig:quantinder}. Several parties on the ``Alice" side, denoted by $A_j$, are to share entanglement with several parties on the ``Bob" side, denoted by $B_k$, such that each $A_j$ is connected to each $B_k$ by an independent channel. Such a network is enabled by ``Charlie", who possesses two identical sources of photon pairs with high-dimensional frequency entanglement. By performing a frequency-resolved entanglement swapping protocol as we describe in this work, Charlie is able to convert the multimode entanglement of the sources into one of many distinct Bell pairs, dependent on his outcome $C_{jk}$, each of which can be routed to a distinct user pair $A_j$ and $B_k$. In this way a single quantum repeater can serve multiple channels, multiplexed in the frequency domain.

In conclusion, we have demonstrated a multimode frequency entanglement swapping scheme that is easily implemented with generic SPDC sources and readily available measurement apparatus. Our design provides a simple way of heralding a high number of orthogonal frequency Bell pairs that is completely measurement-based and requires no source engineering. Alternatively, our protocol could be combined with frequency translators in the signal beams \cite{Wright2017} to generate multiple copies of the same Bell state using broadband sources in a versatile manner, again without the requirement of source engineering. Finally, with the advent of push-button sources of entangled photon pairs \cite{Basset2019}, multiplexed quantum repeaters of the kind that our protocol allows could prove to be a scalable solution for quantum communication networks.

\paragraph*{Acknowledgements.---}\hspace{-3ex}
This project has received funding from the European Union's Horizon 2020 research and innovation programme under Grant Agreement No. 665148, the United Kingdom Defense Science and Technology Laboratory (DSTL) under contract No. DSTLX-100092545, and the National Science Foundation under Grant No. 1620822.

\begin{acknowledgements}
\end{acknowledgements}

\bibliography{biblio}

\onecolumngrid
\pagebreak 
\setcounter{equation}{0}
\renewcommand{\theequation}{Supp.\arabic{equation}}

\section{Supplemental Materials}

\subsection{Four photon state}

\begin{figure}[t]
	\centering
	\includegraphics[width=.35\linewidth]{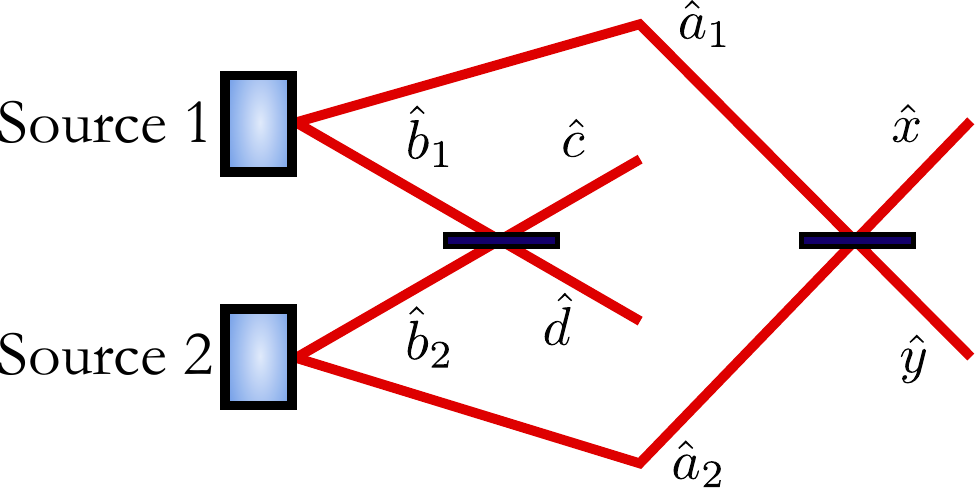}
	\caption{General scheme.}
	\label{fig:supp:simple:scheme}
\end{figure}

To correctly describe the four photon state in our entanglement swapping setup, we begin by labeling the paths with bosinic operators according to Fig\ref{fig:supp:simple:scheme}. Each source generates two pairs of photons by spontaneous parametric down conversion whose signal and idler paths are respectively labelled $\hat{a}_n$ and $\hat{b}_n$, where $n \in \{1,2\}$ denotes the source number. These follow the standard bosonic commutation rules. 
The Hamiltonian describing source $n$ is generally written as:
\begin{align}
	\hat{H}_n = \sqrt{\eta_n} \int \ud \w_s \ud\w_i \ u_n(\w_s+\w_i) \ \textrm{sinc}\left[ \frac{\Delta k_n(\w_s,\w_i) L}{2}\right] \ \hat{a}_n^\dagger(\w_s) \hat{b}_n^\dagger(\w_i) + \textrm{h.c.}
	\label{eq:supp:greeneggsandHamiltonian}
\end{align}
where $u_n$ represents the spectral mode of the pump, $\Delta k$ is the wave-vector mismatch between the pump, signal and idler waves and $\eta_n$ is the gain of the parametric process, which depends on the crystal length $L$, the non-linear strength of the material and the number of photon in the pump beam. 

In the low gain regime, it is straightforward to compute the state at the output of the $n^\textrm{th}$ source:
\begin{align}
	\ket{\psi_n}=\sum_{k=0}^\infty \frac{\sqrt{\eta_n}^k}{k!}\left(\int \ud\omega_s \ud\omega_i f_n(\omega_s, \omega_i) \adag_n(\omega_s) \bdag_n(\omega_i)\right)^k\ket{\mathrm{vac}}
\end{align}
The function $f$ is the joint spectral amplitude (JSA) which defines the energy conservation between the daughter photons. In general, the two sources can be different, but for the sake of generality, we'll assume that they are equivalent, thus having an equal JSA and effective non-linearity.

To derive the state $\hat{\rho}_{RB}$ heralded by the BSM, we begin by writing the SPDC state due to two independent and identical sources as a tensor product
\begin{align}
	\ket{\psi_\textrm{SPDC}} &= \ket{\psi_1}\otimes \ket{\psi_2} \nonumber \\
	&=\left\{\hat{1} + \sqrt{\eta} \int \ud\omega_s \ud\omega_i f(\omega_s,\omega_i)\adag_1(\omega_s)\bdag_1(\omega_i) + \frac{\eta}{2}\left(\int \ud\omega_s \ud\omega_i f(\omega_s,\omega_i)\adag_1(\omega_s)\bdag_1(\omega_i)\right)^2 + \dots \right\}\nonumber\\
	\otimes&\left\{\hat{1} + \sqrt{\eta} \int \ud\omega_s \ud\omega_i f(\omega_s,\omega_i)\adag_2(\omega_s)\bdag_2(\omega_i) + \frac{\eta}{2}\left(\int \ud\omega_s \ud\omega_i f(\omega_s,\omega_i)\adag_2(\omega_s)\bdag_2(\omega_i)\right)^2 + \dots \right\}\vac. \label{eq:SPDCstate}
\end{align}
We expand this and keep only terms of order $\eta$, which are responsible for the four-photon contribution:
\begin{align}
	\ket{\psi_{\mathrm{\eta}}}&\simeq \int \ud\omega_s \ud\omega_i \ud\omega_s' \ud\omega_i' f(\omega_s,\omega_i)f(\omega_s',\omega_i') \adag_1(\omega_s) \bdag_1(\omega_i)  \adag_2(\omega_s') \bdag_2(\omega_i')\vac \nonumber \\
	&+\frac{1}{2}\int \ud\omega_s \ud\omega_i \ud\omega_s' \ud\omega_i' f(\omega_s,\omega_i)f(\omega_s',\omega_i') \adag_1(\omega_s) \bdag_1(\omega_i)  \adag_1(\omega_s') \bdag_1(\omega_i')\vac \nonumber \\
	&+\frac{1}{2}\int \ud\omega_s \ud\omega_i \ud\omega_s' \ud\omega_i' f(\omega_s,\omega_i)f(\omega_s',\omega_i') \adag_2(\omega_s) \bdag_2(\omega_i)  \adag_2(\omega_s') \bdag_2(\omega_i')\vac.
\end{align}
For convenience, we will denote these three terms $\ket{\Psi_{12}}$, $\ket{\Psi_{11}}$, and $\ket{\Psi_{22}}$, so that we have, with the proper renormalization
\begin{equation}
	\ket{\psi_\mathrm{\eta}} = \sqrt{\frac{2}{3}}\left(\ket{\psi_{12}}+\frac{1}{2}\ket{\psi_{11}} + \frac{1}{2}\ket{\psi_{22}}\right), \label{eq:psip}
\end{equation}
and the density matrix for this state is
\begin{equation}
	\hat{\rho}_\mathrm{\eta} = \frac{2}{3}\left(\ket{\psi_{12}}\bra{\psi_{12}} + \frac{1}{4}\ket{\psi_{11}}\bra{\psi_{11}} + \frac{1}{4}\ket{\psi_{22}}\bra{\psi_{22}}\right) +\cancel{ \mathrm{cross\ terms}}.
	\label{rhop}
\end{equation}
The cross terms correspond to coherence between the terms in $\ket{\psi_\eta}$, which is ultimately due to the optical phase of the pump. Because our sources are pumped by the same laser, we do indeed expect them to be mutually coherent. However, over the course of a measurement run (several hours), the phase drifts significantly, so it is reasonable to average over it, and thus these cross terms vanish.
For the rest of the manuscript, we will neglect the terms $\ket{\phi_{11}}$ and $\ket{\phi_{22}}$ from our computations and focus solely on $\ket{\phi_{12}}$.

Most of the experiments that we performed rely on performing a spectrally-resolved Bell state measurement between the idler photon. There are two cases that we should consider, whether a spectral coincidence between the idlers projects the signal into a pure state or into a mixed state.

\subsection{Pure state approximation}

\subsubsection{Heralded state and JSI}

A Bell state measurement is performed on the idler photons by interfering them at a beamsplitter and detecting coincidences at the output while monitoring the frequency of the interfering idler photons. The beamsplitter operation is defined by the following operators:
\begin{equation}
	\cdag(\omega) = \frac{\bdag_1(\omega) + \bdag_2(\omega)}{\sqrt{2}},\quad \dddag(\omega') = \frac{\bdag_1(\omega') - \bdag_2(\omega')}{\sqrt{2}},
\end{equation}
and coincidences are detected between $\hat{c}$ and $\hat{d}$. We consider that we have a perfect resolution in that spectral measurement, such that the POVM element for this detection is simply given by
\begin{align}
	\hat{\Pi}_{jk}^\textrm{BSM} = \cdag(\W_j)\dddag(\W_k)\ket{\textrm{vac}}\bra{\textrm{vac}} \hat{c}(\W_j)\hat{d}(\W_k),
	\label{eq:supp:piBSMpure}
\end{align}
which is a projector onto the monochromatic frequencies $\W_j$ and $\W_k$.

\begin{figure}[t]
	\centering
	\includegraphics[width=.8\linewidth]{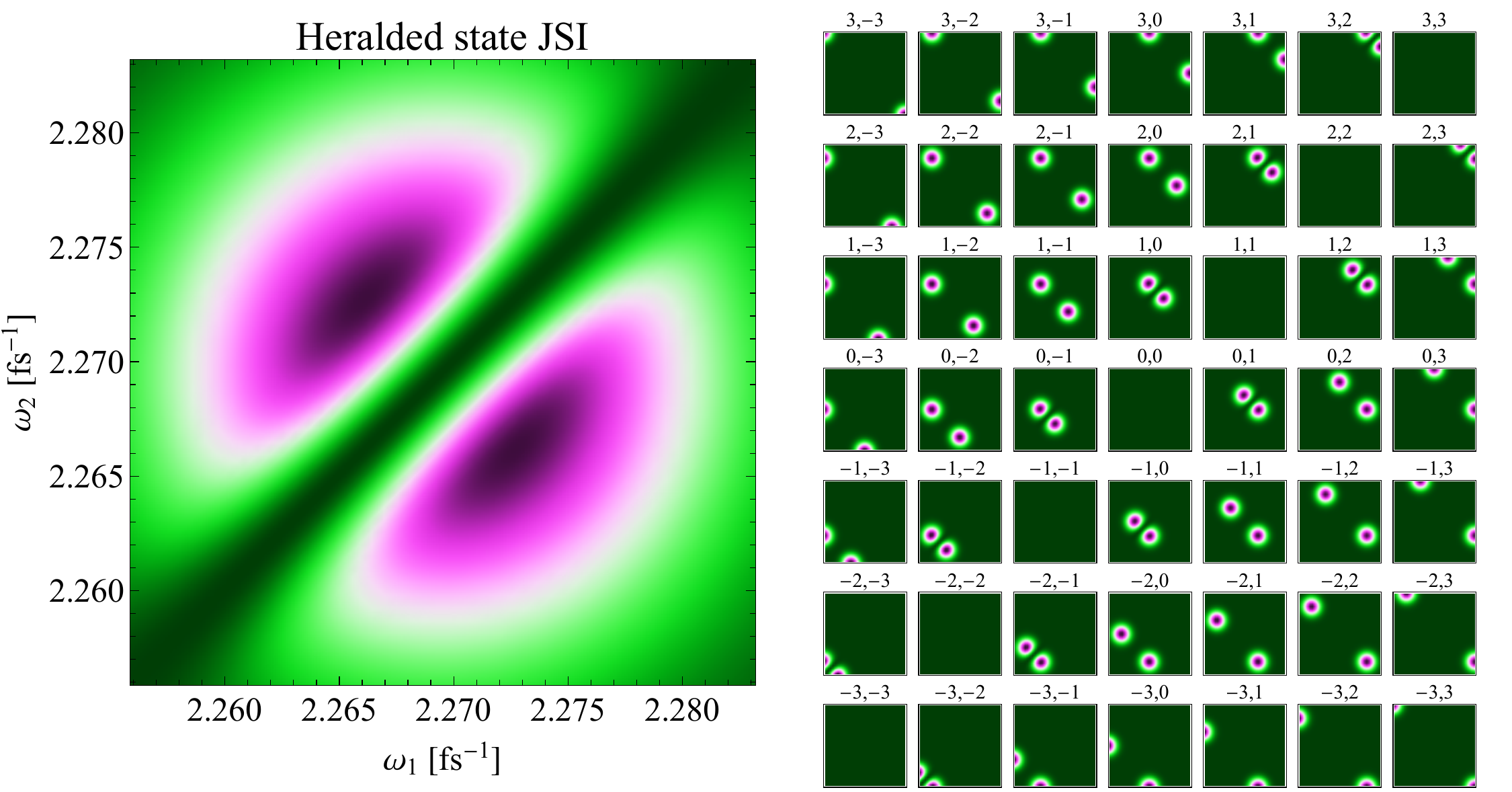}
	\caption{Left: full heralded JSI defined by the sum of Eq.\eqref{eq:supp:heraldedJSIsum} over all $j,k$. Right: $F_{jk}$ for different values labeled as $\lambda_j$ and $\lambda_k$. All the plots on this figures have the same axes.}
	\label{fig:supp:sim:JSI}
\end{figure}

We then proceed to compute the heralded signal state. It is defined by:
\begin{align}
	\ket{\Psi_{jk}^{-}}=\frac{\hat{\Pi}_{jk}^\textrm{BSM}\ket{\psi_{12}}}{\sqrt{p_{jk}}},
	\label{eq:supp:heraldedstatedeff}
\end{align}
where the norm $p_{jk}$ is given by
\begin{align}
	p_{jk}=\bra{\psi_{12}}\hat{\Pi}_{jk}^\textrm{BSM}\ket{\psi_{12}}.
\end{align}
This the probability density of a coincidence between the idler photons at $(\W_j,\W_k)$, or equivalently, the JSA of the idlers \textit{after} the beamsplitter. Upon computing $p_{jk}$, we obtain
\begin{align}
	p_{jk}=\frac{1}{2}\Big[ \rho(\Omega_j,\Omega_j)\rho(\Omega_k,\Omega_k) - \rho(\Omega_j,\Omega_k)\rho(\Omega_k,\Omega_j) \Big],
	\label{eq:supp:pjkpure}
\end{align}
where 
\begin{align}
	\rho(\Omega,\Omega') = \int \ud \w \, f(\w,\Omega) f^\ast(\w,\Omega')
	\label{eq:supp:rhojk}
\end{align}
is the idlers' density matrix. From Eq.\eqref{eq:supp:heraldedstatedeff}, we obtain the following expression for the heralded state:
\begin{align}
	\ket{\Psi_{jk}^{-}}=\frac{1}{\sqrt{\mathcal{C}_{jk}}}
	\frac{\ket{\phi_j}_1\ket{\phi_k}_2 - \ket{\phi_k}_1\ket{\phi_j}}{\sqrt{2}},
	\label{eq:supp:heraldedstate}
\end{align}
where we defined the the functions $\phi_{j(k)}(\omega) = f(\omega,\W_{j(k)}) / \rho(\W_{j(k)},\W_{j(k)})$ and the states $\ket{\phi_{j(k)}}$ as
\begin{equation}
	\ket{\phi_{j(k)}}_{1(2)} = \int \ud \omega \, \phi_{j(k)}(\omega) \adag_{1(2)}(\omega)\vac.
\end{equation}
The normalization of Eq. \eqref{eq:supp:heraldedstate} is given by $\mathcal{C}_{jk} = 1 - \modsqr{\braket{\phi_j | \phi_k}}$ for any heralded frequencies $\W_j$ and $\W_k$. The $\phi_{j(k)}$ functions are normalized but not orthogonal. Their definition follows from the fact that when the idlers frequency is infinitely resolved, the signals are heralded is a pure state. It is therefore convenient to approximate the JSA $f$ as a Gaussian distribution (for instance by approximating the Sinc function by a Gaussian of the same width), such that
\begin{equation}
	f(\w_s,\w_i) = C \, \exp\left[ 
	-\left(\frac{\w_s-\w_0}{2\sigma_s}\right)^2 	-\left(\frac{\w_i-\w_0}{2\sigma_i}\right)^2 	-\alpha (\w_s-\w_0)(\w_i-\w_0),
	\right] \label{eq:supp:JSAsource}
\end{equation}
where $\sigma_{s}$ ($\sigma_i$) is the spectral width projected on the $\w_s$ ($\w_i$) axis, $\w_0$ is the center frequency, $\alpha$ quantifies the amount of spectral entanglement and $C=\left( \int \ud^2 \w \left| f(\w,\w') \right|^2 \right)^{-1/2}$ is a normalization constant. From there, the expression of $\phi_{j(k)}$ is given by
\begin{equation}
	\phi_{j(k)}(\omega) = \frac{1}{\sqrt{\sigma_s\sqrt{2\pi}}}\exp[-(\omega-\omega_{j(k)})^2/4\sigma_s^2],
\end{equation}
where the center frequency of this heralded marginal dependent on a coincidence with an idler at frequency $\W_{j(k)}$ is 
\begin{align}
	\omega_{j(k)}=2\alpha \ \sigma_s^2 \ \W_{j(k)}.
	\label{eq:supp:omegajk}
\end{align}
Under this prescription, any energy variation due to the idler's frequency is absorbed in the normalization, hence to retrieve any quantity that depends on the JSA, a proper weight has to be applied.\\
The heralded joint spectrum is then easily obtained by computing
\begin{align}
	F_{jk}(\w_1,\w_2) &= \left|  \Braket{\w_1, \w_2 | \Psi_{jk}^{-}} \right|^2 \nonumber\\
	&= \frac{1}{2 \mathcal{C}_{jk}} \Big| \phi_j(\w_1)\phi_k(\w_2) - \phi_j(\w_2)\phi_k(\w_1) \Big|^2. \label{eq:supp:heraldedJSIpure}
\end{align}
The heralded JSI is zero when $j=k$, i.e. when the heralding idler photons are indistinguishable. The photons bunch at either output of the idler beamsplitter, and therefore the probability to measure four-fold coincidences is null. Each heralded color of the JSI consists of two identical separable Gaussian joint spectra centered at $\w_j$ and $\w_k$ which separation depends on the heralding frequency $\W_j, \W_k$. A representation of these JSI is given in Fig\ref{fig:supp:sim:JSI}(b), using the Gaussian model which parameters are fitted to the experimental distribution Eq. \eqref{eq:supp:JSAsource} of our source (see Fig\ref{fig:supp:JSI}).

In the absence of spectral resolution in the BSM, the measurement operator becomes

\begin{equation}
	\hat{\Pi}^\text{BSM}=\sum_{j,k} \hat{\Pi}^\text{BSM}_{jk},
\end{equation}
and the heralded state is mixed, given by

\begin{equation}
	\hat{\rho}= \sum_{jk} p_{jk} \ket{\Psi^-_{jk}}\bra{\Psi^-_{jk}}.
\end{equation}
The JSI due to this mixed state is then given by

\begin{align}
	F(\w_1,\w_2) &= \text{Tr}\left(\hat{\rho}\ket{\w_1, \w_2}\bra{\w_1,\w_2}\right)= \sum_{jk} p_{jk}\left|  \Braket{\w_1, \w_2 | \Psi_{jk}^{-}} \right|^2 \nonumber \\
	&= \sum_{jk} p_{jk} F_{jk}(\w_1,\w_2),
	\label{eq:supp:heraldedJSIsum}
\end{align}
which we represented in Fig\ref{fig:supp:sim:JSI}(a). Upon performing the calculation, we find that the expression of the JSI is given by:
\begin{align}
	F(\w_1,\w_2) = \frac{1}{2} \Big[ \rho_s(\w_1,\w_1) \rho_s(\w_2,\w_2) - \modsqr{\rho_s(\w_1,\w_2)} \Big],
	\label{eq:fulljsi}
\end{align}
where $\rho_s(\w,\w') = \int \ud \W f(\w,\W) f^\ast(\w',\W)$ is the signals' density matrix. The expression is similar to \eqref{eq:supp:pjkpure}. Indeed, equation \eqref{eq:fulljsi} is exactly what would be obtained were the beamsplitter placed in the signal paths, rather than the idler paths, and this equivalence is ultimately due to the non-separability of the JSA $f(\omega,
\Omega)$.
\subsubsection{Entanglement verification}

To verify that the heralded state is indeed entangled, we combine both signal photon with another beamsplitter, and label its output according to Fig\ref{fig:supp:simple:scheme}:
\begin{equation}
	\xdag(\omega) = \frac{\adag_1(\omega)e^{i\w \tau} + \adag_2(\omega)}{\sqrt{2}},\quad \ydag(\omega') = \frac{\adag_1(\omega')e^{i\w' \tau} - \adag_2(\omega')}{\sqrt{2}}
\end{equation}
where we introduced a relative delay $\tau$ between the two inputs. We apply this transformation to the heralded state \eqref{eq:supp:heraldedstate} and define our verification POVM as a coincidence between the output of the signal beamsplitter:
\begin{align}
	\hat{\Pi}_\textrm{verif} = \int \ud\w \ud \w' \, \xdag(\w)\ydag(\w')\ket{\textrm{vac}}\bra{\textrm{vac}} \hat{x}(\w)\hat{y}(\w').
\end{align}
The probability of getting a coincidence heralded by a BSM at frequencies $\W_j$ ,$\W_k$ then given by:
\begin{align}
	P_{jk}(\tau) &= \Braket{\Psi_{jk}^-| \hat{\Pi}_\textrm{verif} | \Psi_{jk}^-}
	\label{eq:supp:PjkPure}\\
	&= \int \ud\w\ud\w' \, \Big |\braket{\w , \w' | \Psi_{jk}^-} \Big|^2 \nonumber
\end{align}
where, $\ket{\omega,\omega'} = \hat{x}^\dagger(\omega)\hat{y}^\dagger(\omega')\vac$. As for the heralded JSI, the full probability summed over all possible heralding frequency bins is
\begin{align}
	P(\tau) = \sum_{jk} p_{jk} P_{jk}(\tau).
	\label{eq:fullptau}
\end{align}
Upon evaluating Eq.\eqref{eq:supp:PjkPure}, we obtain
\begin{align}
	P_{jk}(\tau) &= \frac{1}{2} \left(\frac{1 + e^{-\sigma_s^2 \tau^2} \cos{\big[ (\w_j-\w_k) \tau \big]} -|\braket{\phi_j|\phi_k}|^2(1+e^{-\sigma_s^2 \tau^2})}{1-|\braket{\phi_j|\phi_k}|^2}\right), \label{eq:supp:PjkPure2}
\end{align}
\begin{figure}[t]
	\centering
	\includegraphics[width=.8\linewidth]{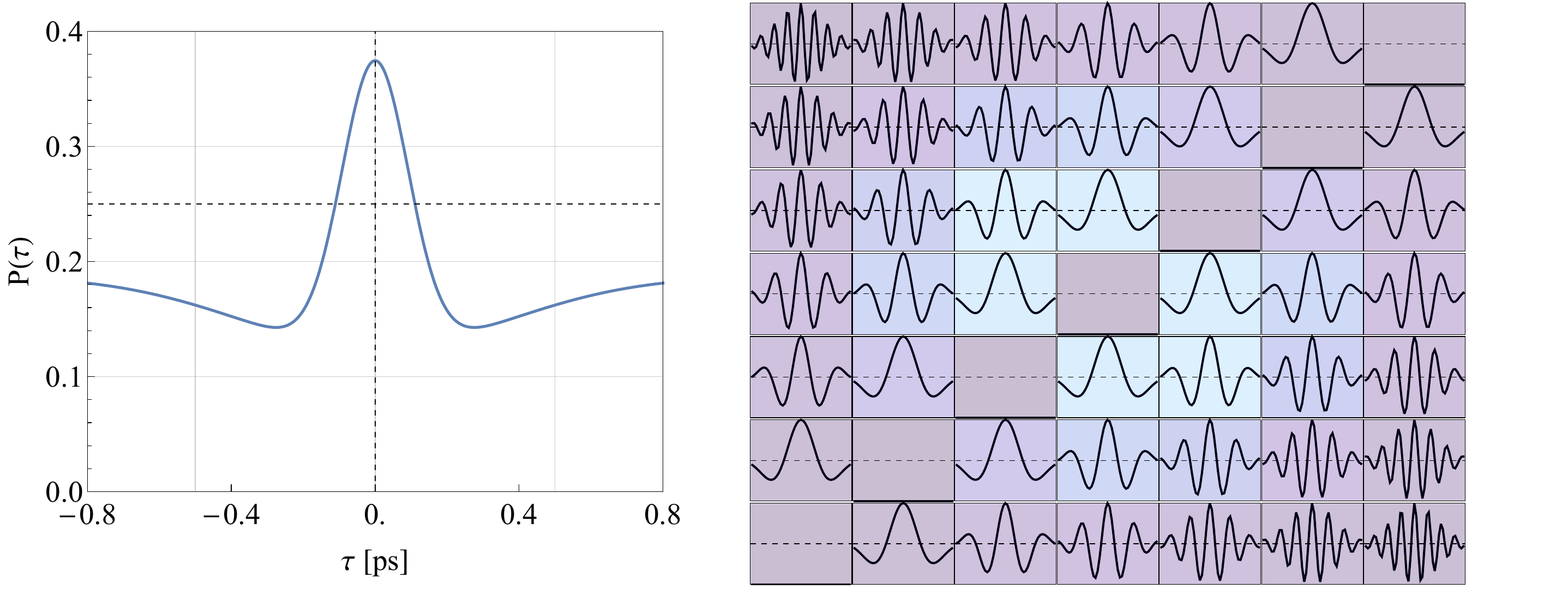}
	\caption{Left: probability $P(\tau)$ of coincidence heralded by a BSM without spectral resolution from Eq.\eqref{eq:supp:peakfull}. Right: $P_{jk}$ for different values labeled as $\lambda_j$ and $\lambda_k$, Eq.\eqref{eq:supp:PjkPure}. The color gradient indicates the probability of getting a coincidence on the herald photons. All the plots on this figures have the same axes.}
	\label{fig:supp:sim:fringes}
\end{figure}
%

which oscillate at the difference frequency $\w_j - \w_k$ between the heralded states. These fringes are a witness of entanglement swapping, which can be simply demonstrated by setting $\alpha \rightarrow 0$ in $\w_{j(k)}$ (see Eq.\eqref{eq:supp:omegajk}), thus removing the oscillating term in Eq.\eqref{eq:supp:PjkPure2}. Therefore, non entangled state will only manifest as a constant term as a function of $\tau$. 

This probability depends on the overlap integral $\braket{\phi_j | \phi_k}$ which quantifies the overlap between the marginals of the heralded state. This quantity goes to zero while the heralding bins $\W_j,\W_k$ are further apart while it goes to unit when they become degenerate. However, asymptotic analysis of Eq.\eqref{eq:fullptau} shows that the weight factor $p_{jk}$ as defined in Eq.\eqref{eq:supp:pjkpure} constrains the total probability to be zero. Therefore, it is reasonable to approximate the spectrally-resolved probability as 
\begin{align}
	P_{jk}(\tau) &\approx \frac{1}{2} \left(
	1 + e^{-\sigma_s^2 \tau^2} \cos{\big[ (\w_j-\w_k) \tau \big]} 
	\right), \label{eq:supp:PjkPureApprox}
\end{align}
to which we fit our experimental results. 
Finally, we can evaluate \eqref{eq:fullptau} without any approximations to obtain
\begin{align}
	P(\tau)=\frac{1}{4}\left(1+\left|\int\ud\w\ud\W \, f^2(\w,\W)e^{i\w\tau}\right|^2-\int \ud^2\W \, \rho(\W,\W')\rho(\W',\W) - \int\ud^2\w \, \rho_s(\w,\w')\rho_s(\w',\w)e^{i(\w-\w')\tau}\right),
	\label{eq:supp:peakfull}
\end{align}
which also depends on the signals' density matrix as in \eqref{eq:fulljsi}. The entanglement verification signal therefore contains four terms. The first one is simply background, while the second one is the overlap integral between the two sources. Evaluating this term while scanning the delay $\tau$ reveals a peak (which is Gaussian in our approximated model) which width depends on the joint temporal distribution of the sources. The last two terms consist respectively on the overlap integrals between the idlers' and the signals' density matrices of each source. The latter consists of an unheralded HOM dip between the signals photons. Hence, the full verification signal corresponds to peak centered in a HOM dip. In Fig\ref{fig:supp:sim:fringes}), we plotted a simulation of the full signal $P(\tau)$ as well as the spectrally resolved probabilities $P_{jk}(\tau)$ for different heralding frequencies $\W_j,\W_k$.

\subsection{Mixed state model}

\subsubsection{Heralded state and JSI}

In the realistic case, the idlers' BSM is not performed with perfect resolution, but rather over a certain spectral window. In our case, this is due to the resolution of the time-of-flight spectrometer, which is a convolution of multiple response function in the frequency-to-time conversion. It is dominated by the timing jitter ($\simeq 30$ ps) of the superconducting nanowires.\\
When this resolution is not perfect, then we can show that the signal photons are heralded into a mixed state. We begin by rewritting the idlers' POVM as:
\begin{equation}
	\hat{\Pi}_{jk}^\mathrm{BSM} = \int \ud\W \ud\W' |t_j(\W)|^2 |t_k(\W')|^2 \cdag(\W)\dddag(\W')\vac\bra{\mathrm{vac}}\hat{c}(\W)\hat{d}(\W')
\end{equation}
where $t_{j(k)}(\W)$ is a filter transmission amplitude centered at $\W_{j(k)}$. It is straightforward to show that the POVM \eqref{eq:supp:piBSMpure} is obtained by setting $t_{j(k)}(\W) \rightarrow \delta (\W-\W_{j(k)})$.\\
It can be seen from the Fig\ref{fig:supp:JSI} that filtering the idler with a finite filter function results in a joint spectrum that is not necessarily separable, and therefore the signal photons are heralded in a mixed state defined by the density matrix:
\begin{align}
	\hat{\rho}_{jk} = \frac{\operatorname{Tr}_{\hat{b}} \Big[ \hat{\Pi}_{jk}^\mathrm{BSM} \ket{\psi_{12}} \bra{\psi_{12}} \Big]}{ \operatorname{Tr} \Big[ \hat{\Pi}_{jk}^\mathrm{BSM} \ket{\psi_{12}} \bra{\psi_{12}} \Big] }
\end{align}
where $\operatorname{Tr}_{\hat{b}}$ is the partial trace over the subspace defined by operators $\hat{b}_1$ and $\hat{b}_2$.\\
Similar to the pure state case, the $p_{jk}$ are defined as
\begin{align}
	p_{jk} &= \operatorname{Tr} \Big[ \hat{\Pi}_{jk}^\mathrm{BSM} \ket{\psi_{12}} \bra{\psi_{12}} \Big] \nonumber\\
	&= \frac{1}{2} \int \ud \W \ud \W' \, |t_j(\W)|^2 |t_k(\W')|^2 \Big( \rho(\W,\W) \rho(\W',\W') - \rho(\W,\W') \rho(\W',\W) \Big)
\end{align}
where the idlers' density matrix is defined as Eq.\eqref{eq:supp:rhojk}, and we again obtain Eq.\eqref{eq:supp:pjkpure} by setting the filters $t_{j(k)}$ as $\delta$ functions.\\
We may now compute the heralded state density matrix:
\begin{equation}
	\hat{\rho}_{jk} =  \int \ud\omega \ud\omega' \ud\tilde{\omega} \ud \tilde{\omega}' \rho_{jk}(\omega,\omega';\tilde{\omega},\tilde{\omega}')\ket{\omega}\ket{\tilde{\omega}}\bra{\omega'}\bra{\tilde{\omega}'} 
	\label{eq:rhojk}
\end{equation}
where
\begin{equation}
	\begin{gathered}
		\rho_{jk}(\omega,\omega';\tilde{\omega},\tilde{\omega}') =\\
		\frac{1}{2p_{jk}}\int \ud\Omega \ud\Omega' |t_j(\Omega)|^2 |t_k(\Omega')|^2 \big(f(\omega,\Omega)f(\tilde{\omega},\Omega')-f(\omega,\Omega')f(\tilde{\omega},\Omega)\big)\big(f^*(\omega',\Omega)f^*(\tilde{\omega}',\Omega')-f^*(\omega',\Omega')f^*(\tilde{\omega}',\Omega)\big)
		\label{eq:supp:rhoheraldedstate}
	\end{gathered}
\end{equation}
The heralded JSI is then given by by:
\begin{align}
	F_{jk}(\w_1,\w_2) = \bra{\w_1,\w_2} \hat{\rho}_{jk} \ket{\w_1,\w_2}
\end{align}
where $\ket{\w_1,\w_2} = \adag_1(\w_1) \adag_2(\w_2) \ket{\textrm{vac}}$. Finally, as in the pure state case, in the absence of frequency resolution at the BSM, the heralded mixed state is 

\begin{align}
	\hat{\rho} = \sum_{jk} p_{jk}\ \hat{\rho}_{jk},
\end{align}
and the JSI for this state is again given by

\begin{align}
	F(\w_1,\w_2)= \sum_{jk} p_{jk} F_{jk}(\w_1,\w_2).
\end{align}

\subsubsection{Entanglement verification}

When the signal photons in the state $\hat{\rho}_{jk}$  are incident on a 50:50 beamsplitter, the coincidence fringes at the output are given by

\begin{align}
	P_{jk}(\tau) = \text{Tr} \left(\hat{\Pi}_\text{verif}\hat{\rho}_{jk}\right)=\int \ud^2\w\bra{\w,\w'}\hat{\rho}_{jk}\ket{\w,\w'},
\end{align}
where $\ket{\w,\w'}=\xdag(\w)\ydag(\w')\vac$ as before. When evaluated, this gives
\begin{align}
	P_{jk}(\tau) = \frac{1}{2p_{jk}} \int \ud\Omega \ud\Omega' |t_j(\Omega)|^2 |t_k(\Omega')|^2 \Big( \rho(\W,\W) \rho(\W',\W') - \rho(\W,\W') \rho(\W',\W) \Big) P(\W,\W',\tau),
\end{align}
where
\begin{align}
	P(\W,\W',\tau)=\frac{1}{2}\left(\frac{1+e^{-\sigma_s^2\tau^2}\cos{[2\alpha\sigma_s^2(\W,\W')\tau]} - \mathcal{O}(\W,\W')(1-e^{-\sigma_s^2\tau^2})}{1-\mathcal{O}(\W,\W')}\right),\quad \mathcal{O}(\W,\W') = \frac{\rho(\W,\W')\rho(\W',\W)}{\rho(\W,\W)\rho(\W',\W')},
\end{align}
is the pure state interference expression from \eqref{eq:supp:PjkPure2}.\\

Since, by construction, $\sum_{j,k} |t_j(\W)|^2 |t_k(\W')|^2 = 1$, the integrated coincidence probability is, as before,

\begin{equation}
	\begin{gathered}
		P(\tau)=\sum_{j,k} p_{jk} P_{jk}(\tau)=\frac{1}{4}\int \ud\Omega \ud\Omega'\Big( \rho(\W,\W) \rho(\W',\W') - \rho(\W,\W') \rho(\W',\W) \Big) P(\W,\W',\tau),
	\end{gathered}
\end{equation}
which is equivalent to \eqref{eq:supp:peakfull}.

\subsection{Detection}
The single photon are detected utilizing superconducting nanowire single photon detector (SNSPD) from IDQuantique (ID281) which can detect the arrival time of photons with a resolution of 20 ps. This temporal resolution is translated into spectral resolution using time-of-flight spectrometers (TOFS), thanks to frequency-to-time conversion \cite{torres2011,goda2013}. For coarse spectral resolution, we used two  spools of 500 meters-long HP780 fiber. These imprint a dispersion of about 50 ps/nm, hence the spectrometers have a resolution of 0.4 nm. The losses per spool at 830 nm are about 33\%. For fine resolution, we used two chirped fiber Bragg gratings (CFBG from Teraxion) with a dispersion of 1000 ns/nm \cite{davis2017}, or a spectral resolution of less than 0.02 nm. This extra resolution comes with a heavy loss of over 80\% and also with a finite spectral window of 10 nm. The signals coming out of the detectors are registered with a time-to-digital converter (TDC, ID900 from IDQuantique). The time reference is provided by the clock generated by the laser source, thus ensuring that each time tag is taken with respect to a stable signal for each pulse. With this setup, it is possible to register coincidences between any combination of the four photons with the advantage of measuring their wavelength. This allows for ``pixelization'' of any event into spectral bins.

To ensure that our TOFS are accurate, it is necessary to calibrate them. This procedure is usually realized using single frequency emission from known white light sources, but this isn't possible with TOFS since they require a pulsed signal to extract their time tags. While it is usually sufficient to use rough estimate of the dispersion imprinted by the fiber spool or by the CFBG, this doesn't take into account any other source of dispersion in the setup. Therefore, we opted for an in-situ calibration utilizing the single photons from the SPDC.

We utilized a pulse shaper based on putting a spatial light modulator (SLM) in a 4-f line, enabling to address both the amplitude and phase over a 30 nm range with a resolution of 0.02 nm. By scanning a narrow interference filter of 1 nm FWHM over the SLM mask while recording the resulting time tags, we obtain a linear dependency between the recorded time tags and the wavelength of the filter set on the pulse shaper. The slope of that function is then the dispersion parameter of the TOFS. With the CFBG-based spectrometers, we obtained a dispersion of $944\pm  4$ ps/nm and $946\pm  2$ ps/nm, respectively. With the fiber spools, we measured an equal dispersion of $-54\pm  1$ ps/nm. With the calibration in hand, we may now link time of arrival of each photon to their wavelength in any configuration shown in Fig\ref{fig:setup}.

\subsection{Entangled photon source}

\subsubsection{Source distinguishability}

The entanglement verification protocol we use, that is, the two-photon interference of the state $\ket{\Psi^-_{jk}}$, ultimately relies on the indistinguishability of the two source states. To see this, we relabel the source JSA's as $f_1(\w,\W)$ and $f_2(\w,\W)$, and for simplicity, we assume that they are identical up to a translation in frequency space. Note now that this leads to a heralded state

\begin{align}
	\ket{\Psi^-_{jk}} \propto \ket{\phi^1_j}_1\ket{\phi^2_k}_2-\ket{\phi^1_k}_1\ket{\phi^2_j}_2,
\end{align}
where
\begin{align}
	\ket{\phi^{1(2)}_{j(k)}} = \int \ud \w \phi^{1(2)}_{j(k)}(\w)\adag_{1(2)}(\w)\vac,
\end{align}
and
\begin{align}
	\phi^{1(2)}_{j(k)} = \frac{f_{1(2)}(\w,\W_{j(k)})}{\rho_{1(2)}(\W_{j(k)},\W_{j(k)})}.
\end{align}

\begin{figure}[t]
	\centering
	\includegraphics[width=.5\linewidth]{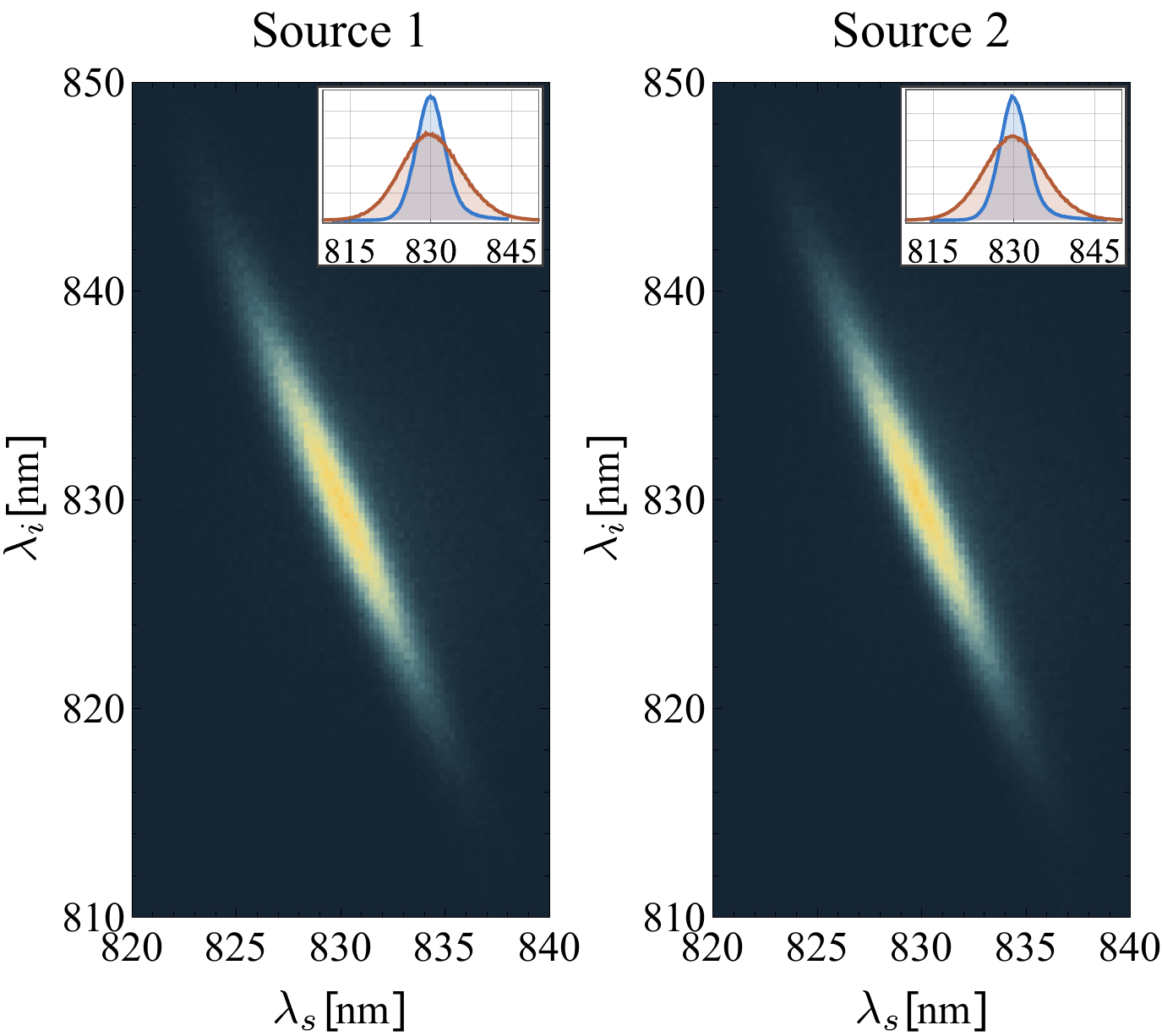}
	\caption{Joint spectral intensity of both sources. Insets: marginal spectra.}
	\label{fig:supp:JSI}
\end{figure}

Although this state is still entangled, this entanglement cannot in general be assessed through measuring coincidence fringes in $P_{jk}(\tau)$, because the distinguishability of $f_1$ and $f_2$ will reduce the visibility of these fringes. To see this, we recalculate $P_jk(\tau)$ in its approximate form \eqref{eq:supp:PjkPureApprox}, and find

\begin{align}
	P_{jk}(\tau) \approx \frac{1}{2}\left(1-V_{jk}e^{\sigma_s^2\tau^2}\cos{\left[\left(\frac{\w^1_j+\w^2_j}{2}-\frac{\w^1_k+\w^2_k}{2}\right)\tau\right]}\right)
\end{align}
where the visibility $V_{jk}$ is given by

\begin{align}
	V_{jk}=\left(\int\ud\w\phi^{1*}_j(\w)\phi^2_j(\w)\right)\left(\int\ud\w\phi^{1*}_k(\w)\phi^2_k(\w)\right).
\end{align}

We can maximize this visibility by maximizing the overlap $f_1$ and $f_2$. We see that this latter provides a lower bound on $V_{jk}$ by writing

\begin{align}
	\int\ud\w\phi^{1*}_j(\w)\phi^2_j(\w)=\frac{\int\ud\w f_1^*(\w,\W_j)f_2(\w,\W_j)}{\sqrt{\rho_1(\W_j,\W_j)\rho_2(\W_j,\W_j)}}\geq \int\ud\w\ud\W f_1^*(\w,\W)f_2(\w,\W),
\end{align}
and likewise for $k$.

It is relatively straightforward to maximize the quantity on the left by tuning experimental parameters, namely pump wavelength, phasematching angle, and transverse optical fiber position (due to residual spatial chirp), and observing two-fold coincidences resulting from first order interference of the sources. Because both sources are pumped with the same pulse, the two-photon term of the state is given by

\begin{align}
	\ket{\psi}\propto \int \ud\w\ud\W\left(f_1(\w,\W)\adag_1(\w)\bdag_1(\W) + f_2(\w,\W)\adag_2(\w)\bdag_2(\W)\right)\vac.
\end{align}
A straightforward calculation shows that the probability of a two-fold coincidence between ports $\hat{c}$ (or $\hat{d})$ and $\hat{x}$ (or $\hat{y}$) is given by

\begin{align}
	P_{cc} = \frac{1}{4}\int \ud^2\w \Big|f_1(\w,\W) \pm f_2(\w,\W)\Big|^2=\frac{1}{2}\Big(1 \pm \mathrm{Re}\int \ud^2\w f_1^*(\w,\W)f_2(\w,\W)\Big) \label{eq:supp:ccfringes}
\end{align}

\begin{figure}[t]
	\centering
	\includegraphics[width=.45\linewidth]{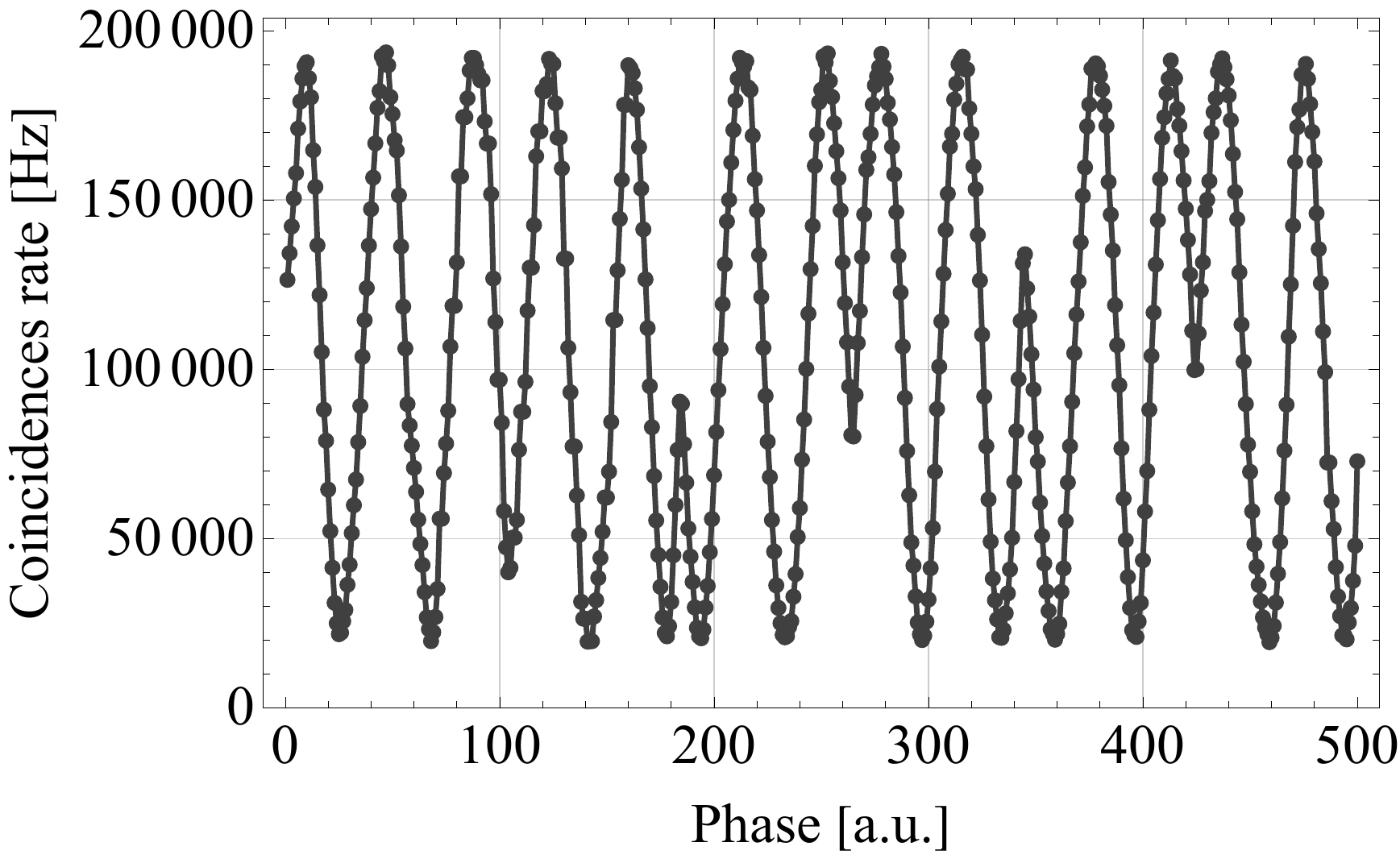}
	\caption{Measured coincidence fringes $P_{\textit{cc}(\tau)}$ with a contrast of $80\%$.}
	\label{fig:supp:coincfringes}
\end{figure}

In the following, we will outline additional measurements to quantify the source indistinguishability. In our case, our dual-pass geometry implies that we need to match the JSD of both sources, which is achieved when both signals and idlers from both sources have maximum overlap. We opted for a bulk crystal source in Type II to enable pumping in both directions while being able to separate our four photons into different paths. We used a BiBO crystal due to its relatively high non linearity.

First, we measured the JSI by directing the two daughter photons from either source into the fiber spools, since their large spectral bandwidth would be cropped with the CFBG. The JSI from each source is depicted in Fig\ref{fig:supp:JSI}. They show that both sources are nearly indistinguishable; a singular value decomposition yields a Schmidt number of $K_1 = 2.9 \pm .1$ and $K_2 = 2.9 \pm .1$. These values are lower than the theoretical expectation ($K\sim 5$) because of the timing jitter of our detectors that result in a wider distribution. This was confirmed by measuring the JSI with the CFBG that have a better resolution but are limited in range. The correlation width was found to be lower and therefore the Schmidt number can be expected to be at least $K=4$.

Note that this method is insensitive to any spectral phase difference, such as dispersion from the pump, since the second pump is slightly more dispersed than the first due to propagation. This has been shown to increase the entanglement and the Schmidt number \cite{Davis2020,Ansari2018}. However, this difference should be negligable, and the method presented latter that relies on Eq.\eqref{eq:supp:ccfringes} allows for a more accurate estimation of the overlap. Nevertheless, the JSI measurement showed perfect correspondance between the intensity of the two sources which is a critical step to ensure indistinguishably between the uncorrelated photon pairs.

To further characterize the indistinguishability of the sources, we measure their heralded $g^{(2)}$ by splitting their signal photon into a beamsplitter. This yields a value of $g_1^{(2)} = 0.16 \pm 0.003$ and $g_2^{(2)} = 0.14 \pm 0.003$. These values are consistent with the relatively high optical power that is utilized to pump the sources in order to maximize the probability of four-fold coincidences. The lower value of $g^{(2)}$ for source 2 is consistent with the fact that it also has a higher heralding efficiency than source 1. The reason is not entirely clear, but it is likely that the previous interaction with the PDC crystal on the first pass results in an additional filtering on the pump as well as a slight reduction in optical power because of absorption.

Finally, in Fig\ref{fig:supp:coincfringes} we measured the coincidences between ports $\hat{c}$ and $\hat{x}$ (see Fig\ref{fig:supp:simple:scheme}) while scanning the relative phase between the two pump fields with a piezoelectric stack, which is related to the probability from Eq.\eqref{eq:supp:ccfringes}. We scanned using a slow voltage ramp resulting in a few micrometers of displacement over a few seconds. The visibility of those fringes is of $80\%$, which is a direct measurement of the overlap between the two sources, and therefore a quantification of distinguishability.\\
Note that we also performed this measurement with spectral resolution, essentially measuring those interferences in narrower spectral bins using our time-of-flight spectrometer. This results in interferences over a much narrower bandwidth, therefore restricting the degrees of freedom of the single photon mode-function. Notably this method has the advantage of being less sensitive to higher order phase mismatch between both sources, such as dispersion. The measured contrast across all spectral bins was found to be $90\%$.

\begin{figure}[t]
	\centering
	\includegraphics[width=.95\linewidth]{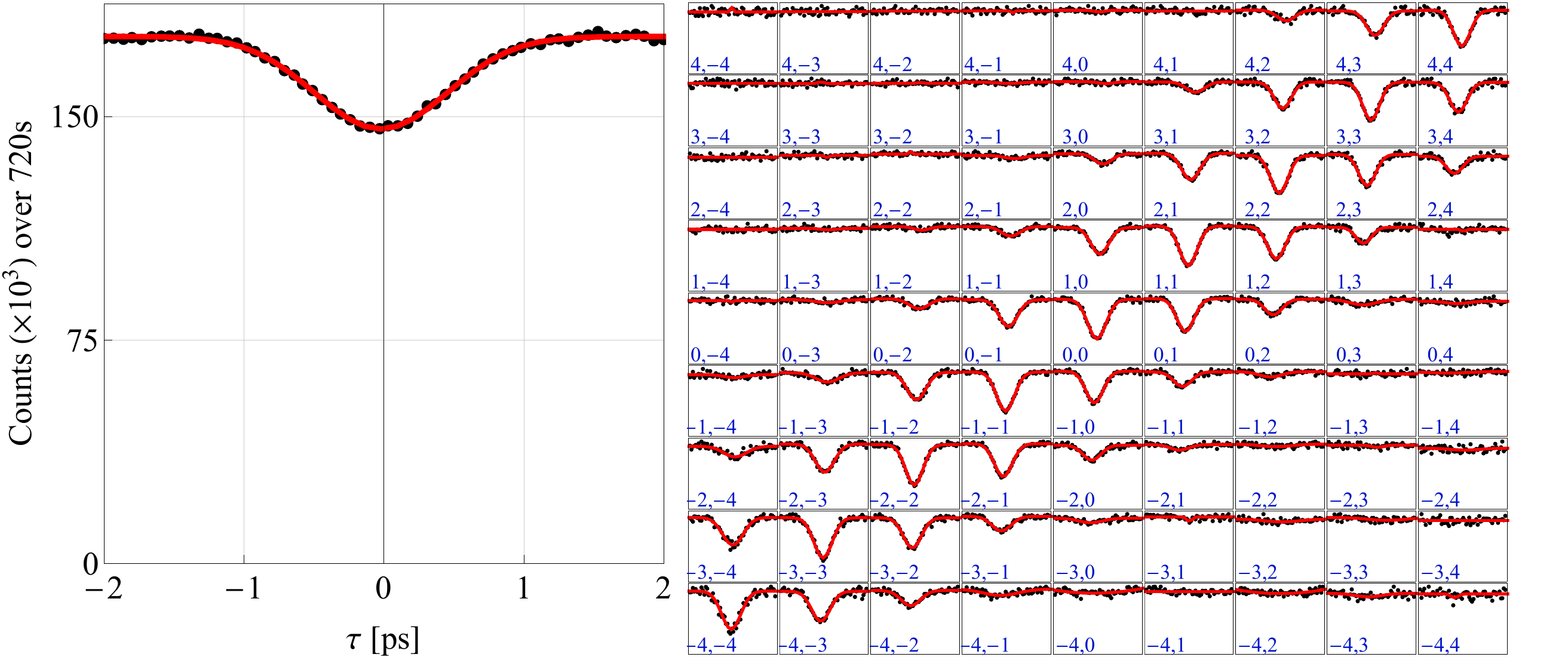}
	\caption{Left: HOM dip between the signal photons heralded by a coincidence between the idler photons. Right: same measurement but with spectral resolution of the heralding photons, labelled $j,k$ for $\W_k,\W_k$, where index $j,k=0$ corresponds to the center frequency $\w_0$.}
	\label{fig:supp:homdip}
\end{figure}

\subsubsection{Purity of the heralded states}

Since the state $\ket{\psi_{12}}$ from the sources is assumed to be a pure state, the purity of the heralded states $\ket{\Psi^-_{jk}}$ is ultimately dependent on the amount of spectral filtering in the heralding BSM. To assess this purity, we measure HOM interference between the heralded signal photons when there is no beamsplitter in the idler arms. In this case, upon a coincidence detection of the idler photons at $(\Omega_j,\Omega_k)$, the reduced state of the signal photons is separable, and given by

\begin{align}
	\hat{\rho}_j\otimes\hat{\rho}_k = \left(\int \ud^2\omega \rho_j(\omega,\omega')\right)\left(\int \ud^2\tilde{\omega}\rho_k(\tilde{\omega},\tilde{\omega}')\right)\adag_1(\omega)\adag_2(\tilde{\omega})\ket{\text{vac}}\bra{\text{vac}}\hat{a}_1(\omega')\hat{a}_2(\tilde{\omega}'),
\end{align}
where

\begin{align}
	\rho_{j(k)}(\omega,\omega') = \int \ud\Omega |t_{j(k)}(\Omega)|^2 f(\omega,\Omega)f^*(\omega',\Omega).
\end{align}

When the signal photons in this state are incident on a 50:50 beamsplitter, the expected visibility of the HOM interference is given by \cite{Mosley2008}

\begin{align}
	V=\mathrm{Tr}(\hat{\rho}_j\hat{\rho}_k),
\end{align}
and when the idlers are detected in identical frequency bins $(j=k)$, this becomes

\begin{align}
	V=\mathrm{Tr}(\hat{\rho}_{j(k)}^2)=\mathcal{P}(\hat{\rho}_{j(k)}),
\end{align}
where $\mathcal{P}(\cdot)$ denotes the purity of a state. Thus, for $(j=k)$ the visibility of the HOM dip gives a lower bound on the purity of the state $\hat{\rho}_{j(k)}$, and by extension, the state $\hat{\rho}_{jk}$. Our measurements, shown in Fig.~\ref{fig:supp:homdip}, indicate that purity of the heralded states is \textit{at least} 70\%, as evidenced by the HOM visibility along the $j=k$ line. By comparison, a direct calculation of the expected purity using our experimental parameters gives $\sim 78\%$. The purity of our heralded state seems to be dominated by the spectral resolution of our spectrometer. Without spectral resolution, the purity of the heralded state is about 20\% as shown in Fig.~\ref{fig:supp:homdip}.

\subsubsection{Background signal}

As shown by Eq.\eqref{rhop}, the full four photon state in the interferometer (see Fig.~\ref{fig:supp:simple:scheme}) contains a contribution from photon pairs emitted by individual sources due to the stochastic nature of parametric down conversion. These terms contribute to $P(\tau)$ in the form of interferences that get averaged over the course of a measurement. It is therefore possible to remove that contribution from the signal subsequently to the measurement by blocking a source and recording the rate of four-fold coincidences.

\begin{figure}[t]
	\centering
	\includegraphics[width=\linewidth]{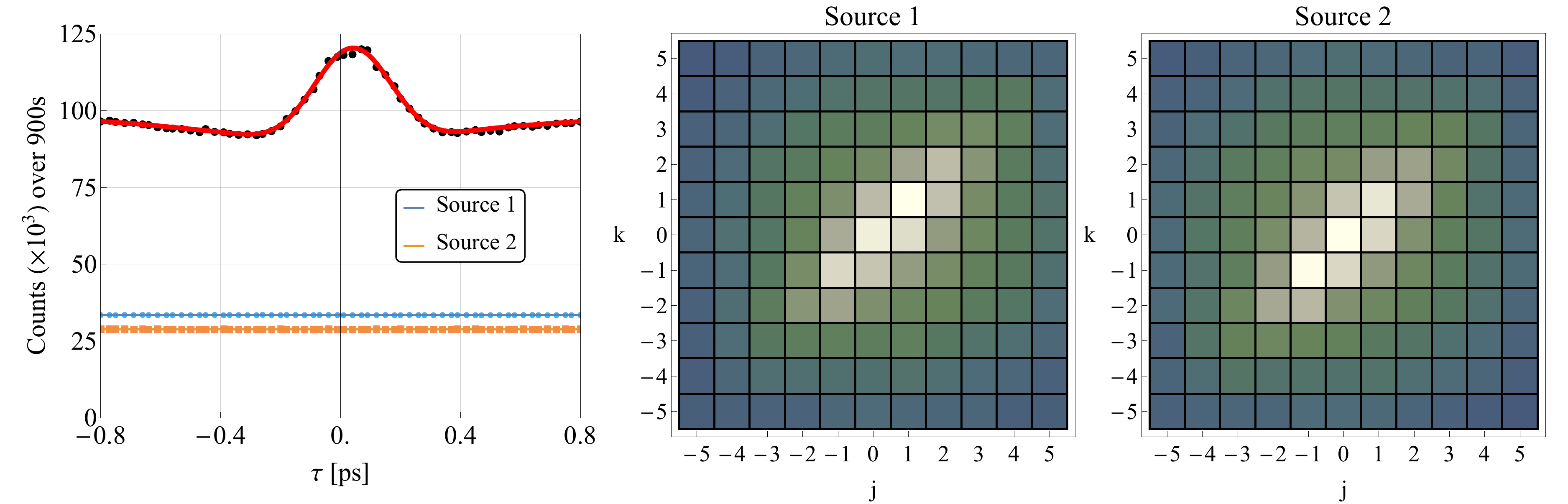}
	\caption{Left: $P(\tau)$ without removing the constant two photon contribution from source 1 (dot) and source 2 (square). $P(\tau)$ ressembles Eq.\eqref{eq:supp:peakfull} and the fit is obtained by summing the individual fits of $P_{jk}$ as given by \eqref{eq:supp:PjkPureApprox}. Right: distribution of these background terms as a function of the heralding frequencies $\W_j,\W_k$}
	\label{fig:supp:background}
\end{figure}

We therefore repeated the measurement of $P_{jk}(\tau)$ with either source blocked to obtain the constant background signal for each $j,k$ frequencies, as shown in Fig.~\ref{fig:supp:background}. This shown that the background terms are similar between both sources, therefore the two sources are similar. Summing over all the bins, we can plot on the same scale the contribution of all term in Fig.~\ref{fig:supp:background}. The peak corresponds to interferences from $\ket{\psi_{12}}$ while the flat terms represent $\ket{\psi_{11}}$ and $\ket{\psi_{22}}$. As expected from the theory, both source contribute to $1/4$ of the full signal. Removing those backgrounds at $\W_j,\W_k$ from $P_{jk}$, we obtain the fringes from the main paper with optimal visibility.

\subsubsection{Orthogonal modes}

From Eq.\eqref{eq:supp:heraldedstate}, we see that the heralded state $\ket{\Psi_{jk}}$ is dependent on the modes $\ket{\phi_j}$ and $\ket{\phi_k}$, which, in the pure state case, results in a heralded joint spectrum (Eq.\eqref{eq:supp:heraldedJSIpure}) dependent on the outer products $\phi_j(\w_1)\phi_k(\w_2)$. For each heralding bin $j$ and $k$, we label the heralded JSI from \eqref{eq:supp:heraldedstate} as $F_n(\w_1,\w_2)$, where $n$ indexes a pair $(j,k)$. These are normalized as $\int \ud^2\w F_{n}(\w_1,\w_2) = 1 ~ \forall ~ n$ but are not orthogonal, even in the pure state case, \textit{i.e} 
$\int \ud^2\w F_n(\w_1,\w_2)F_m(\w_1,\w_2)  \neq \delta_{nm}$. Orthogonality is usually a corner stone in any quantum protocol, and it is therefore necessary to select the heralded states from our measurement that are orthogonal. To do so, we utilize our measurement of $F_{jk}$, obtained by measuring the spectral coincidences between the signal's photon heralded by a BSM on the idlers. We then obtain a figure similar to Fig.~\ref{fig:supp:sim:JSI} albeit without perfect spectral resolution, putting us in the mixed state configuration, but the strategy to select orthogonal modes within this set is similar to the pure state model.

\begin{figure}[t]
	\centering
	\includegraphics[width=\linewidth]{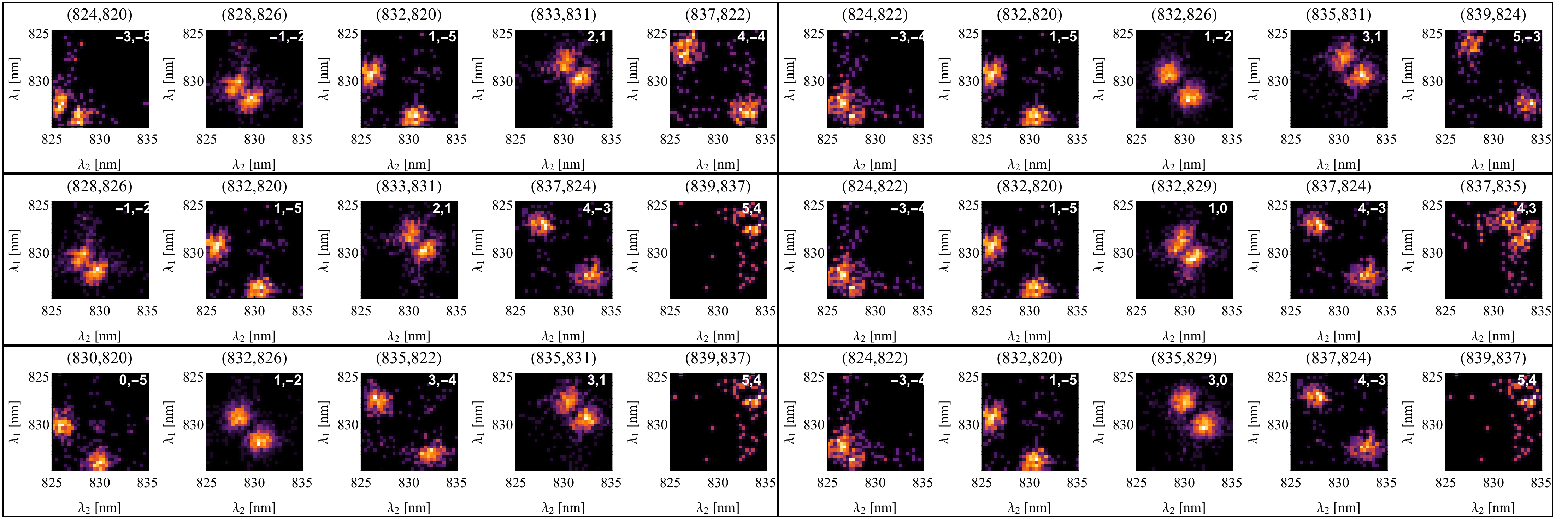}
	\caption{Sets of orthogonal modes $F_{jk}$ that have less than 15 \% of mutual overlap within each part of the grid. The insets label $j$ and $k$.}
	\label{fig:supp:orthomode}
\end{figure}

First, it is important to notice the symmetry in \eqref{eq:supp:heraldedstate}, where $F_{jk} = F_{kj}$. Since our TOFS are well-calibrated, it is reasonnable to symmetrize our measured heralded JSI by averaging the experimentally obtained $F_{jk}$ and $F_{kj}$ (for $j\neq k$) thus defining the $F_n$ functions. Then we compute the mutual overlaps $ \int \ud^2\w F_n(\w_1,\w_2)F_m(\w_1,\w_2) $ and use an algorithm to select a set of modes $\left\{F_{n}\right\}$ which all have an overlap below a certain threshold of $15\%$. We represented a few of these JSI in Fig.~\ref{fig:supp:orthomode}. Since the spectral range of our high resolution TOFS is limited, so is the range over which we can compute overlap, as can be seen from the modes that are labelled with a large $j,k$. Nevertheless, there is a sufficient amount of spectral coincidence in those cases to infer orthogonality with the other JSI.

Note that while this overlap is computed between the joint spectral intensities and not between the states, it can be shown that if the overlap in intensity is zero, then the states are necessarily orthogonal, hence the strategy is valid to select which $\ket{\Psi_{jk}}$ are mutually orthogonal. Therefore, it is fair to say that the JSI $F_{jk}$ from Fig.~\ref{fig:supp:orthomode} correspond to the heralded states $\ket{\Psi_{jk}}$ that are all mutually orthogonal.

\end{document}